\documentclass[floatfix, preprint, showpacs, showkeys, preprintnumbers, nofootinbib, superscriptaddress]{revtex4-1}
\usepackage[utf8]{inputenc}
\usepackage[sort&compress]{natbib}
\usepackage{ulem}
\usepackage{bm}
\usepackage{times}
\usepackage{amssymb,amsbsy,amsmath,amsfonts}
\usepackage{graphicx}
\usepackage{float}
\usepackage{color}
\usepackage{morefloats}
\usepackage{rotating}
\usepackage{srcltx}
\usepackage{slashed}
\usepackage{subfigure}
\usepackage{multirow}
\usepackage{verbatim}
\usepackage{hyperref}
\usepackage{tabularx}
\renewcommand{\arraystretch}{1.2}
\newcommand{\RNum}[1]{\uppercase\expandafter{\romannumeral #1\relax}}

\usepackage{diagbox,}

\begin{document}

\title{Nuclear binding energies in artificial neural networks}

\author{Lin-Xing Zeng}
\affiliation{School of Physics,  Beihang University, Beijing 102206, China}

\author{Yu-Ying Yin}
\affiliation{School of Physics,  Beihang University, Beijing 102206, China}

\author{Xiao-Xu Dong}
\affiliation{School of Physics,  Beihang University, Beijing 102206, China}

\author{Li-Sheng Geng}
\email[E-mail: ]{lisheng.geng@buaa.edu.cn}
\affiliation{Peng Huanwu Collaborative Center for Research and Education, Beihang University, Beijing 100191, China}
\affiliation{School of
Physics,  Beihang University, Beijing 102206, China}
\affiliation{Beijing Key Laboratory of Advanced Nuclear Materials and Physics, Beihang University, Beijing 102206, China }
\affiliation{School of Physics and Microelectronics, Zhengzhou University, Zhengzhou, Henan 450001, China }

\begin{abstract}

The binding energy (BE) or mass is one of the most fundamental properties of an atomic nucleus. Precise binding energies are vital inputs for many nuclear physics and nuclear astrophysics studies. However, due to the complexity of atomic nuclei and of the non-perturbative strong interaction, up to now, no conventional physical model can describe  nuclear binding energies with a precision below 0.1 MeV, the accuracy needed by nuclear astrophysical studies. In this work, artificial neural networks (ANNs), the so called ``universal approximators", are used to calculate  nuclear binding energies. We show that the  ANN can describe all the nuclei in AME2020 with a root-mean-square deviation (RMSD) around 0.2 MeV, which is  better than the best macroscopic-microscopic models, such as FRDM and WS4.
The success of the ANN is mainly due to the proper and essential input features we identify, which contain the most relevant physical information, i.e., shell, paring, and isospin-asymmetry effects. We show that the well-trained ANN has excellent extrapolation ability and can  predict binding energies for  those nuclei so far inaccessible  experimentally. In particular, we highlight the important role played by ``feature engineering'' for physical systems where data are relatively scarce, such as nuclear binding energies.

\end{abstract}


\maketitle

\section{Introduction}

The atomic nucleus is a quantum many-body system with an extremely complex structure~\cite{Zelevinsky2015AtomicNA}. As one of the most fundamental properties of  atomic nuclei,  binding energies (BE) can provide crucial  information on nuclear shapes~\cite{deRoubin:2017fdd}, shell effects~\cite{Mo:2014mqa,Rosenbusch:2015yma} , pairing effects~\cite{Lunney:2003zz}, and the disappearance as well as  emergence of magic numbers~\cite{Porter:2022mtw,Rosenbusch:2015yma}. In addition, binding energies are essential inputs for  superheavy nuclei syntheses~\cite{Tanaka:2020abd} and  nuclear astrophysical studies~\cite{Burbidge:1957vc}, e.g., the r-process~(\cite{Boehnlein:2021eym,Mumpower:2015ova}), X-ray bursts~\cite{Schatz:2016fjp}, and etc. Therefore, reliable theoretical predictions and experimental measurements of  nuclear binding energies have always been at the frontier of nuclear physics~\cite{Moller:2015fba,Xia:2017zka,10.1088/1674-1137/ac7b18}.

In the latest atomic mass evaluation (AME 2020)~\cite{Wang:2021xhn}, the masses of 3556 nuclei (including measured and extrapolated) are compiled. However, various theoretical models predict that  about 8000 to 10000 nuclei may exist~\cite{osti_1513818,Geng:2005yu,Moller:2015fba,Xia:2017zka}, including most of those relevant in nuclear elements syntheses.  Therefore, reliable and accurate theoretical predictions are  in urgent need. Some of the most widely used theoretical models include the Weizsacker-Skyrme model (WS)~\cite{Weizsacker:1935bkz,Wang:2010dm,Wang:2010uk,Liu:2011ama,Wang:2014qqa}, the Relativistic Mean Field model (RMF)~\cite{GAMBHIR1990132,Geng:2005yu}, the Duflo-Zuker model (DZ)~\cite{Duflo:1995ep}, the Hartree-Fock-Bogoliubov model~\cite{Goriely:2013nxa, Goriely:2013xba,Goriely:2009zz, Goriely:2016sdz}, the Finite-Range Droplet model  (FRDM)~\cite{Moller:1993ed,moller_new_2012,Moller:2015fba}, 
and the RCHB~\cite{Xia:2017zka} and DRHBc~\cite{DRHBcMassTable:2022uhi} models. Most of these models can describe the experimental data with a root-mean-square deviation~(RMSD) ranging from about 0.3 MeV to several MeV. Among them, FRDM2012~\cite{moller_new_2012} achieved an RMSD of 0.570 MeV while the Weizsacker-Skyrme (WS4)~\cite{Wang:2014qqa} model gives the best description with an RMSD of 0.298 MeV. In general. the macro-micro models,  rather than the more ``physical" microscopic models, perform better in describing nuclear masses because their parameters are determined by fitting to all the (then available) experimental data.

In recent years,  artificial neural networks~(ANNs), as one of the most powerful machine learning methods, have been successfully applied in  nuclear physics studies~\cite{Boehnlein:2021eym,Harris:2022qtm}, e.g., binding energies~\cite{Utama:2015hva, Utama:2017wqe, Utama:2017ytc,Yuksel:2021nae,Niu:2018csp}
, charge radii~\cite{Dong:2021aqg, Dong:2022wkd,Wu:2020bao,Utama:2016tcl}, $\alpha$-decay half-lives~\cite{Li:2022ifg} , $\beta$-decay half-lives~\cite{Niu:2018trk}, and fission fragment yields~\cite{Ma:2020bic, Ma:2020mbd, wang:2019pct}. 

The studies of nuclear binding energies (masses) can be divided into two categories, i.e., either fitting to the experimental data  directly or to the residuals between  experimental data and model predictions.  In Refs.~\cite{Carnini:2020lvr,Gao:2021eva,Niu:2018csp,Zhang:2017zvb,Utama:2015hva,WU2022137394}, mass residuals are utilized to refine the theoretical models. In Refs.~\cite{Utama:2015hva, Utama:2017wqe, Utama:2017ytc,Niu:2018csp}, Bayesian neural networks are found to be able to describe  nuclear binding energies with an RMSD ranging from 0.266 to 0.850 MeV.  The RMSD obtained in the WS4 supplemented with Light Gradient Boosting Machine (LightGBM) is
$0.170\pm0.011$ MeV~\cite{Gao:2021eva}. The Bayesian machine learning (BML) method proposed in Ref.~\cite{niu_nuclear_2022} achieves an RMSD of 84 keV, the first crossing the 100 keV threshold.

However, there are fewer works that study the experimental data directly. 
In Refs.~\cite{Yuksel:2021nae,Bahtiyar:2022wph}, feed-forward neural networks with different structures are explored. Ref.~\cite{Yuksel:2021nae} yielded an RMSD of 1.84 MeV for 1071 nuclei contained in  AME2016~\cite{wang_ame2016_2017} as the test set.~\footnote{The rather poor performance may be attributed to the fact that  the MLP model has only been trained 800 epochs.} Ref.~\cite{Bahtiyar:2022wph} applied the data augmentation technique to expand the data set, and the RMSD decreased to 1.322 MeV for the test set within the training data region and 1.495 MeV for the new nuclei beyond the training data region.
In Refs.~\cite{Mumpower:2022peg,Lovell:2022pkw}, mixed density networks with 12 physically motivated features~\cite{Lovell:2022pkw} or eight features constrained by the GK relation~\cite{Mumpower:2022peg}  are devised to describe nuclear
mass excesses. In the latter work~\cite{Mumpower:2022peg},
  an RMSD of 0.316 MeV for the test set and 0.186 MeV for the training set for nuclei with $Z\geq20$ was achieved, whose performance is comparable to that of WS4~\cite{Wang:2014qqa}.

In this work, we develop  an ANN with seven input features of most relevance. We find that among the 12 features studied in Ref.~\cite{Lovell:2022pkw}, only six of them are effective in our network. Meanwhile, we find that taking GeLU~\cite{hendrycks2016gaussian} as the activation function enhances the predictive power of the ANN.   Our ANN provides a better description of nuclear binding energies than all the conventional models and in addition shows good extrapolation ability.

This article is organized as follows. In section II, we explain how to construct the ANN and determine the physically motivated input features. Results and discussions are presented in Section III. A short summary and outlook is provided in Section IV. 

\section{Theoretical Formalism}
In this section, we  introduce the ANN and mass data  we used in detail.
\subsection{Artificial neural network}

\begin{figure}[htbp]
  \centering
  \includegraphics[width=1\textwidth]{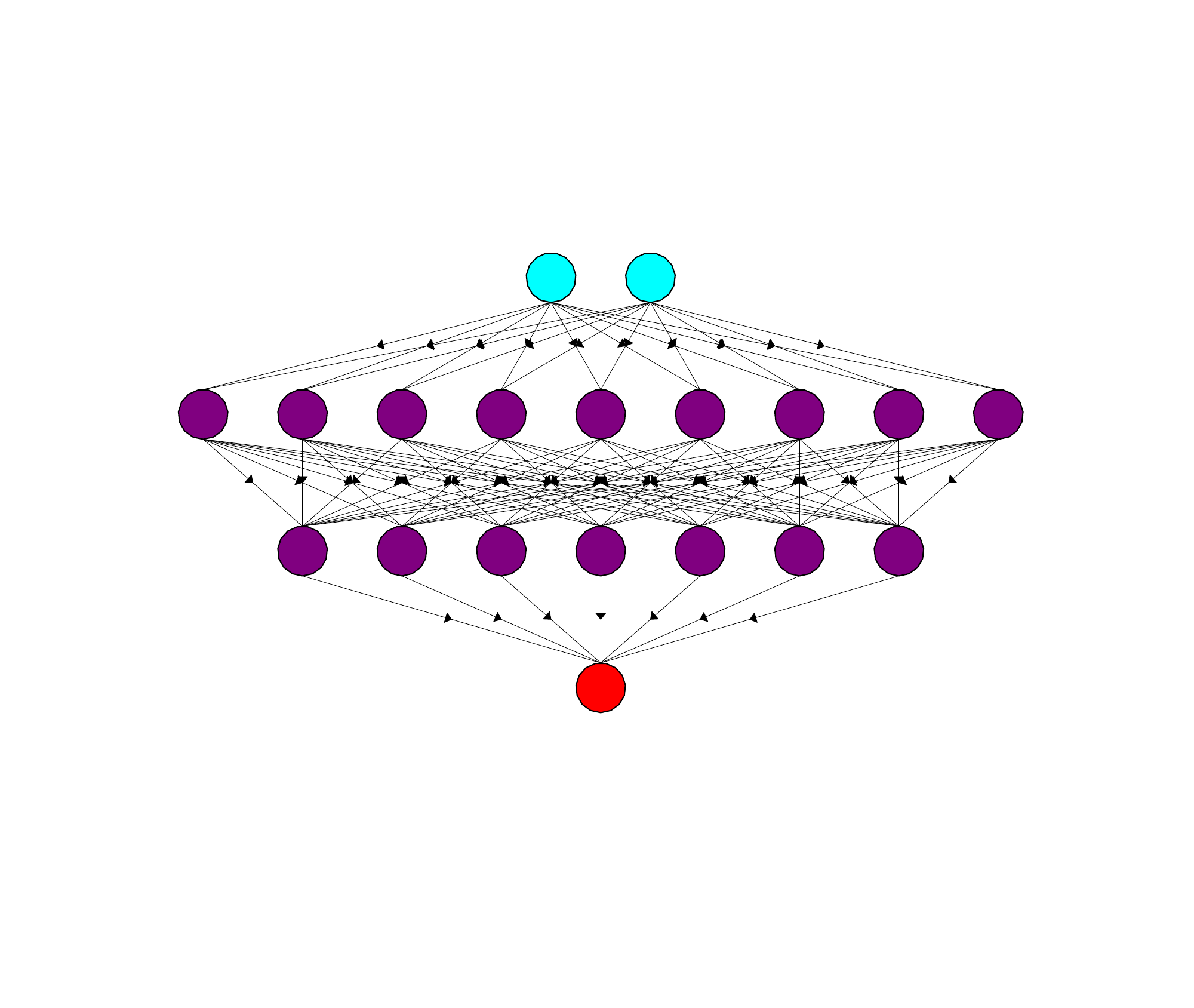}
  \caption{Architecture of a neural network consisting of seven input features, two hidden layers of nine and eight nodes, and one output layer. }\label{pic:neural network}
\end{figure}

Generally speaking, an ANN is a supervised machine learning method which is also regarded as a ``universal approximator". The ANN used in this work is a fully connected feed-forward neural network, consisting of one input layer with seven features,  two hidden layers, and one output layer as shown in Fig.~\ref{pic:neural network}.
The inputs $I_j$ and outputs $O_j$ of layer $j$ are connected as follows
\begin{equation}
O_j=f(W_j\cdot I_j +b_j),
\end{equation}
where $j$ runs over the input layer and the hidden layers, $W_j$ are the weights, $b_j$ are the bias, and $f$ is the activation function to  be specified. For the output layer, no activation function is needed.

Although in principle  one could improve the description of BEs with either more hidden layers or more nodes in each hidden layer, one often ends up with the over-fitting problem. By trial and error, we find that with two hidden layers and about  800 parameters, our ANNs can well describe the binding energies. For a ANN with $I$ inputs, two hidden layers, and one output,  denoted as $[I,H1,H2,O]$, the  number of parameters is $(I+1)\times H1+(H1+1)\times H2+(H2+1)\times O$. Table ~\ref{table:param num} lists the nodes and number of parameters of the different ANNs investigated in the present work. Note that to better understand how the different input features affect the performance of ANNs, in addition to the default ANN with seven features, we also study three other ANNs, where two, four, and six features are used.  For the activation function, we choose GeLU ~\cite{hendrycks2016gaussian},  which is found to perform better than Tanh. For the loss function, we use the standard mean absolute error (MAE):
\begin{equation}\label{eq:loss func}
  LOSS=\frac{\sum^{N}_{i=1}|BE^{th}_{i}-BE^{exp}_{i}|}{N}.
\end{equation}

For numerical implementation, we use the optimized  tensor library PYTORCH~\cite{10.5555/3454287.3455008} and employ the Adam algorithm ~\cite{kingma2014adam} with a learning rate $0.0001$ and the decay constants  $0.9$ and $0.999$. The weight matrices of our ANNs are initialized in PYTORCH with the same random seed.

\begin{table}[htbp]
  \centering
  \caption{Structure, number of parameters, and input features of the different ANNs studied in the present work. }\label{table:param num}
  \begin{tabular}{cccl}
  \hline\hline
    Model & Structure & Number of parameters & Input features \\ \hline
    ANN2 & [2, 35, 19, 1] & 809 & $Z$, $N$ \\
    ANN4 & [4, 35, 17, 1] & 805 & $Z$, $N$, $Z_{EO}$, $N_{EO}$ \\
    ANN6 & [6, 32, 17, 1] & 803 & $Z$, $N$, $Z_{EO}$, $N_{EO}$, $\Delta Z$, $\Delta N$ \\
    ANN7 & [7, 32, 16, 1] & 801 & $Z$, $N$, $Z_{EO}$, $N_{EO}$, $\Delta Z$, $\Delta N$, ASY \\
    \hline\hline
  \end{tabular}
\end{table}

A supervised ANN maps  inputs to the desired outputs. In the present case, the output is the binding energy of a nucleus. As an atomic nucleus is completely determined by its proton and neutron numbers, one can naively take them as  the only inputs.  Nevertheless, it is well known that for small data sets, engineered features (in addition to these ``fundamental features'') , which encode important information (priors) about the system under investigation, can play an invaluable role in enhancing the capacity of ANNs. Such a technique is widely used in nuclear physics studies (see, e.g., ~\cite{Dong:2021aqg, Dong:2022wkd, Mumpower:2022peg, Lovell:2022pkw}). In Ref.~\cite{Dong:2021aqg},  it was shown that in addition to $N$ and $Z$, with two more features accounting for the pairing  and  shell-closure effects, one is able to describe the nuclear charge radii much better than the Bayesian models without these two features. In particular, one is able to describe the strong odd-even staggerings of the charge radii of the calcium and  potassium isotopes. In Ref.~\cite{Dong:2022wkd}, it was shown that the description can be further improved with two more features accounting isospin dependence and local anomalies.

In the studies of binding energies,  in addition to the above mentioned pairing, shell-closure, and isospin dependence effects, many other features have been studied~\cite{Lovell:2022pkw,Li:2022ifg}. In the present work, we find that the most relevant features are those just mentioned, i.e., pairing, shell-closure, and isospin dependence.  The pairing effects are encoded in $Z_{EO}$ and $N_{EO}$, which is 1 when $Z/N$ is  odd, or  0 otherwise.  Shell effects are introduced via $\Delta Z$ and $\Delta N$, which are the differences between $Z$ and $N$ and the closest magic numbers. In this work, the magic numbers are taken to be 8, 20, 28, 50, 82, 126, and 184. As one moves from the beta-stability line, isospin-asymmetry becomes large. Therefore, we take into account  this effect by introducing the seventh feature, ASY, which is defined as 
\begin{equation}\label{eq:ASY}
  ASY=\left( 1-\frac{\kappa}{A^{1/3}}+\xi\frac{2-|I|}{2+|I|A} \right)I^{2}Af_{s}
\end{equation}
where the parameters $ \kappa$, $\xi$, and $f_s$  are taken from WS4~\cite{Wang:2014qqa}.

\subsection{Mass data}

AME2020~\cite{Wang:2021xhn}, in which the masses of 3556 nuclei (including measured and extrapolated ones) are compiled, is referred to as the data set in this work. The training set and test set are extracted from AME2020 as follows. The nuclei in the training set are those which were already in AME2016 ~\cite{wang_ame2016_2017}, and the rest of nuclei that are not in AME2016 but in AME2020 are chosen as the test set (test20).  Based on this selection, there are 3434 nuclei in the training set and 122 nuclei in the test set.

\section{Results and Discussions}
To quantify how well the ANN can describe nuclear binding energies in the training and test sets,
we use the standard root-mean-square deviation (RMSD), $\sigma_{rms}$, defined as
\begin{equation}\label{eq:rms error}
  \sigma_{rms}=\sqrt{ \mathop{\sum}^{N}_{i}\frac{\left(BE^{\mathrm{th}}_{i}-BE^{\mathrm{exp}}_{i}\right)^{2}}{N} },
\end{equation}
where $BE_i^\mathrm{th}$ are the ANN predictions and $BE_i^\mathrm{exp}$ are the experimental binding energies contained in the training and test sets~footnote{In the present work, we do not distinguish between the measured and extrapolated masses compiled in the mass evaluations (AME2016 and AME2020)}.

\renewcommand{\arraystretch}{1}
\setlength\tabcolsep{8pt}
\begin{table}[htbp]
  \centering
  \caption{RMSDs for the training set (consisting of 3434 nuclei) and test set (consisting of 122 nuclei) achieved using different network structures. }\label{table:train16test20}
  \begin{tabular}{c|ccc}
    \hline\hline
    \multirow{2}*{Model} & \multicolumn{3}{c}{$\sigma_{rms}$ (MeV)}  \\
    {} & Training set & Test set & Entire set\\ \hline
    ANN2 & 1.183 & 1.053 & 1.178 \\
    ANN4 & 0.548 & 0.628 & 0.551 \\
    ANN6 & 0.289 & 0.514 & 0.299 \\
    ANN7 & 0.190 & 0.340 & 0.197 \\
    \hline\hline
  \end{tabular}
\end{table}

\begin{table}[htbp]
  \centering
  \caption{Comparisons between  ANN7 and the WS4 model~\cite{WS4_dataset}, for nuclei with $Z\geq8$ and $N\geq8$ compiled in AME2020,  i.e., 3336 nuclei and 120 nuclei contained in our training set and test set, respectively.}\label{tab:ws4}
  \begin{tabular}{c|ccc}
    \hline\hline
    \multirow{2}*{Model} & \multicolumn{3}{c}{$\sigma_{rms}$ (MeV)}  \\
    {} & Training set & Test set & Entire set\\ \hline
    ANN7 & 0.149 & 0.336 & 0.159 \\
    WS4 & 0.415 & 1.295 & 0.474 \\
    \hline\hline
  \end{tabular}
\end{table}

The $\sigma_{rms}$ obtained with different number of input features are shown in Table ~\ref{table:train16test20}. For the entire set, the RMSD reduces from 1.178 MeV in ANN2 to 0.197 MeV in ANN7. This demonstrates unambiguously that using engineered features that explicitly encode the pairing, shell and isospin-asymmetry effects is able to significantly improve the capacity of ANNs to describe/predict nuclear binding energies. We stress that the total number of parameters are similar for all the four network structures. From ANN2 to ANN4, with the features ($Z_{eo}$ and $N_{EO}$), the descriptions improve in not only the training set but also the test set. The RMSD  decreases by almost 50 percent from ANN2 to ANN4.  The explicit consideration of shell effects ($\Delta Z$ and $\Delta N$) further improves the descriptions, and the RMSD for the entire set crosses the $0.3$ MeV threshold. Finally, the explicit consideration of the asymmetry effects further improves the description, and the RMSD of ANN7 (for the training set and entire set) falls  below $0.2$ MeV.

To put the performance of ANN7 into better perspective, we compare it with one of the most refined conventional models, WS4, which only studied those nuclei with $Z\geq8$ and $N\geq8$, i.e., only 3336 nuclei and 120 nuclei among those contained in our training and test sets. The corresponding $\sigma_{rms}$'s are given in Table~\ref{tab:ws4}. Three things are noteworthy. First, ANN7 performs better than WS4 for all the three sets of data. Second, removing the light nuclei with $Z<8$ or $N<8$, the RMSD of ANN7 decreases from 0.2 to 0.16 for the entire set. Third, from the training set to the test set, the RMSDs of both ANN7 and WS4 increase. Somehow surprisingly, in terms of percentage, the increase of WS4 is even larger than that of ANN7. We note in passing that for the 2353 nuclei studied in WS4~\cite{Wang:2014qqa}, ANN7 gives an RMSD of 0.154 MeV, which should be compared with that of WS4, 0.293 MeV.

\begin{figure}[htbp]
  \centering
  \includegraphics[width=15cm]{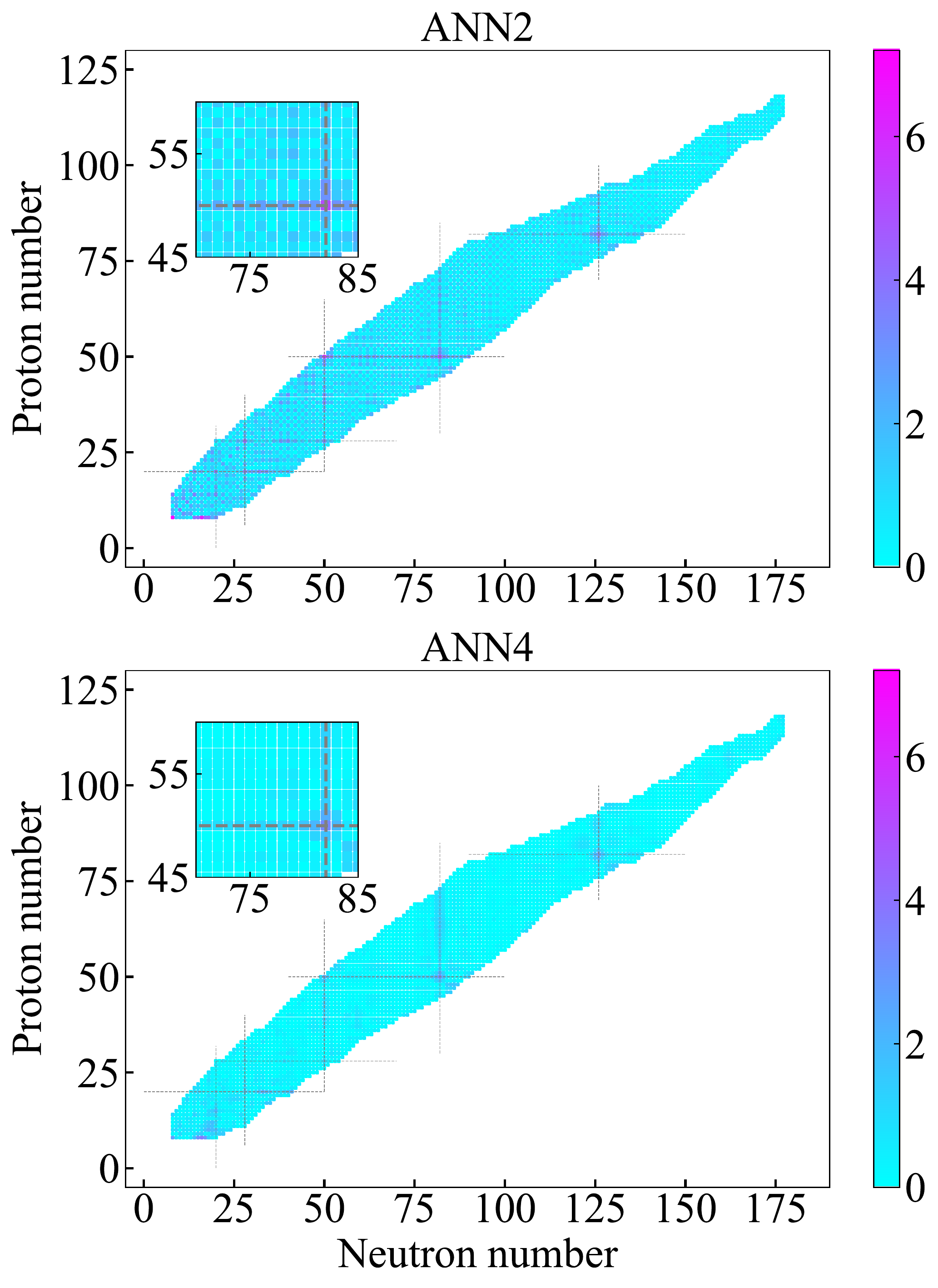}
  \caption{Absolute deviations of  the ANN2 and ANN4 predictions from the experimental  binding energies. The gray lines denote the magic numbers.}\label{pic:ANN-scatter1}
\end{figure}

\begin{figure}[htbp]
  \centering
  \includegraphics[width=15cm]{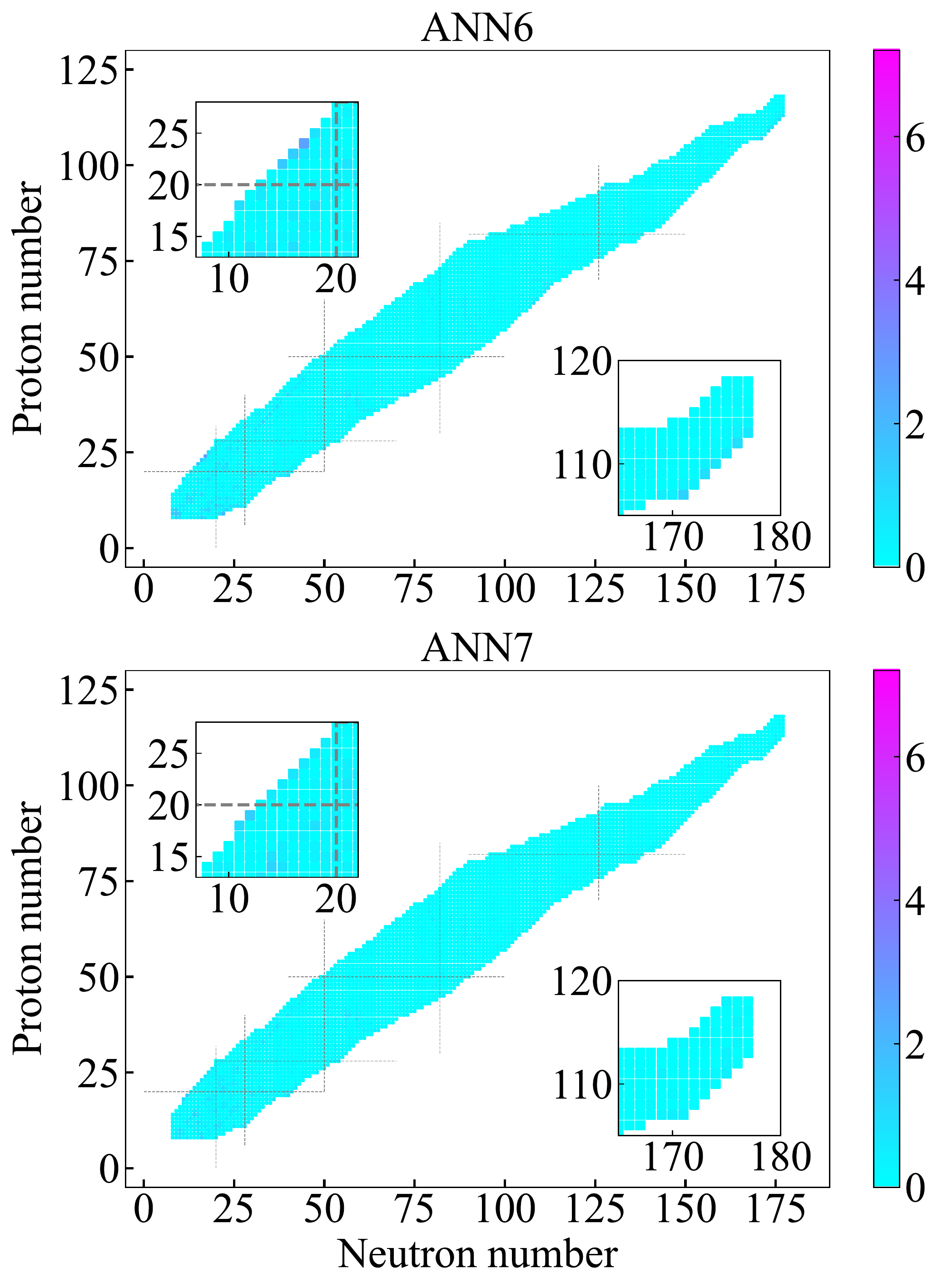}
  \caption{Absolute deviations of  the ANN6 and ANN7 predictions from the experimental  binding energies. The gray lines denote the magic numbers.}\label{pic:ANN-scatter2}
\end{figure}

Fig.~\ref{pic:ANN-scatter1} and Fig.~\ref{pic:ANN-scatter2} provide more details on the deviations of the ANN predictions from the experimental data.
As can be seen from Fig.~\ref{pic:ANN-scatter1}, ANN2 performs relatively worse for even-even nuclei than for their neighboring nuclei (even-odd or odd-odd). With two more  features $Z_{EO}$ and $N_{EO}$, which take into account explicitly the pairing effects, ANN4 improves the description of even-even nuclei, and the $\sigma_{rms}$ is reduced from 1.053 MeV to 0.628 MeV for the test set and from 1.183 MeV to 0.548 MeV for the training set. However, ANN4 does not capture the shell effects. One can see from the bottom panel of Fig.~\ref{pic:ANN-scatter1}, the deviations between the ANN predictions and the experimental data are larger for those nuclei with either proton or neutron number being magic.  For doubly magic nuclei, the deviation is particularly large. With the shell effects taken into account, the ANN6 successfully describes these nuclei as can be seen  from the upper panel of  Fig.~\ref{pic:ANN-scatter2}. A closer examination of Fig.~\ref{pic:ANN-scatter2}  reveals that for heavy nuclei with $N\sim175$ and light nuclei with $N\sim20$ and $Z\sim25$, the deviations are relatively large. We note that these nuclei are more neutron rich which are newly compiled in AME2020
. As a result, one can anticipate that an improved description can be achieved by considering a feature that explicitly takes into account isospin asymmetry. This is indeed the case.  The $\sigma_{rms}$ of ANN7 for the test set decreases from 0.514 MeV to 0.340 MeV, even though the deviations for some light nuclei are still relatively large.

\begin{figure}[htbp]
  \centering
  \includegraphics[width=15cm]{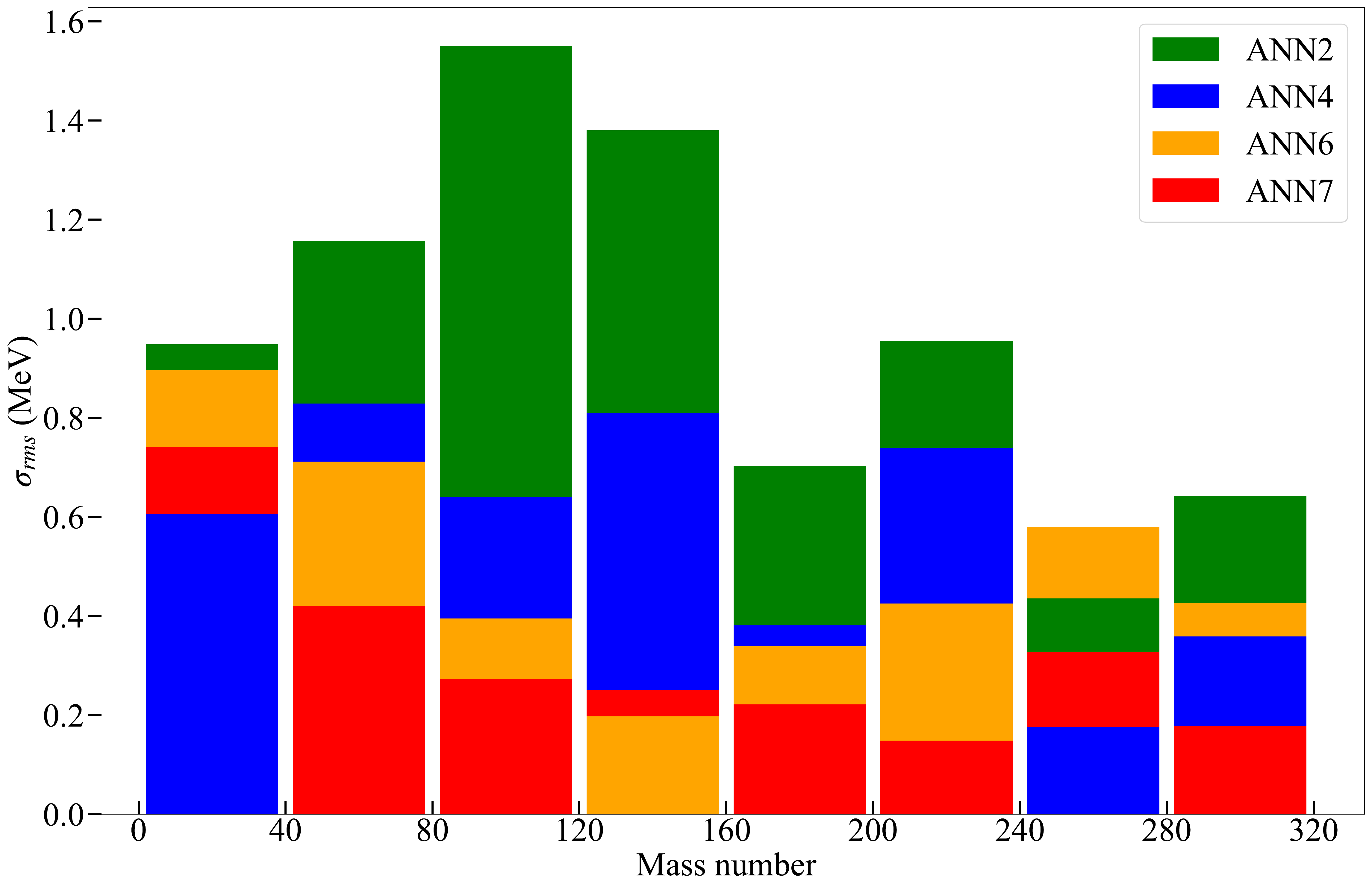}
  \caption{Distributions of $\sigma_{rms}$ between every 40 mass numbers in test20.}\label{pic:ANN-Error-dist-A}
\end{figure}

\begin{figure}[htbp]
  \centering
  \includegraphics[width=16cm]{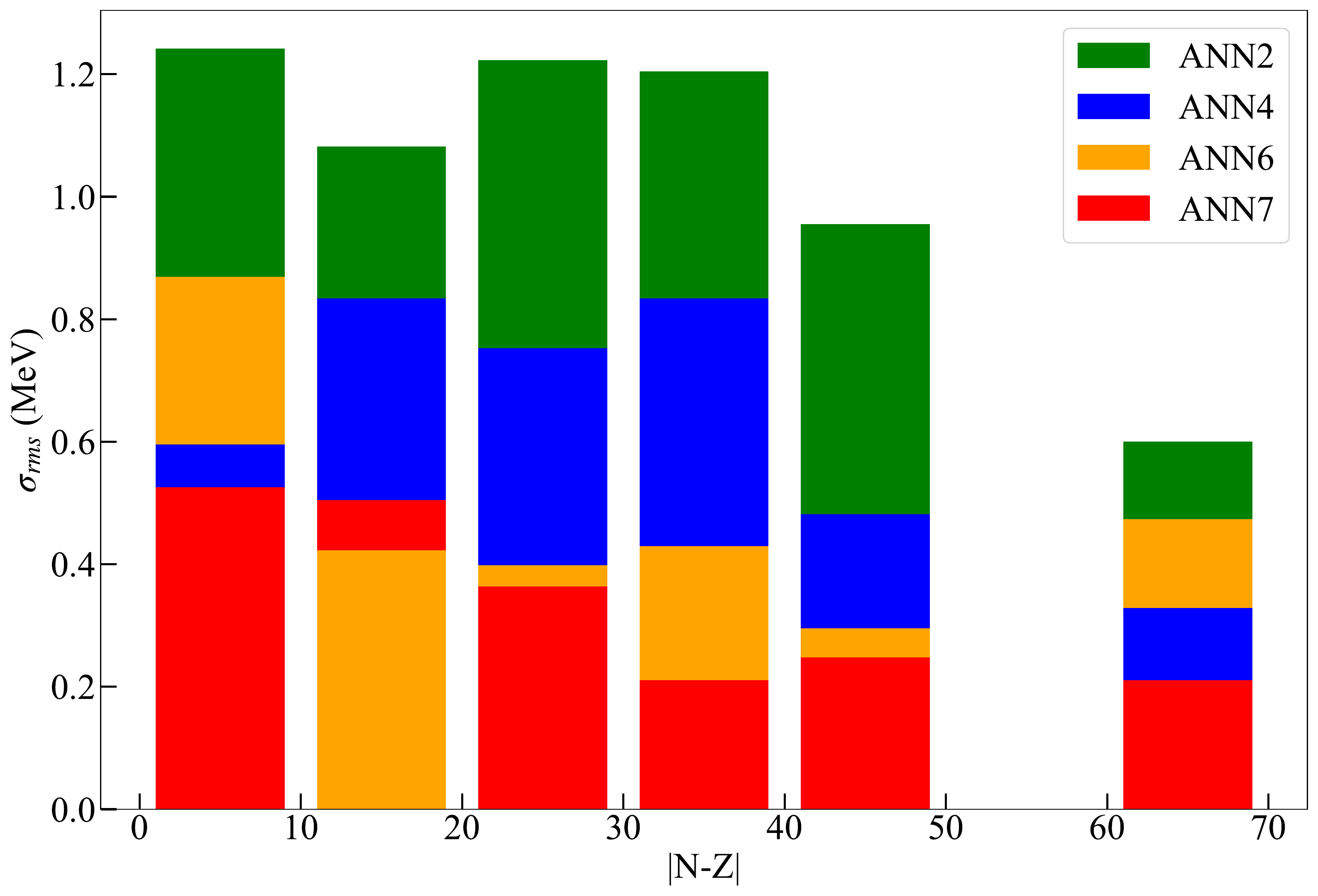}
  \caption{$\sigma_{rms}$ as functions of  $|N-Z|$, which reflects the  ability of ANNs in describing nuclei with large isospin asymmetries. Note that there is no nucleus in test20 with $50<|N-Z|<60$.}\label{pic:ANN-Error-N-Z}
\end{figure}

It is instructive to examine how the different ANNs perform for nuclei in different mass regions. In 
Fig.~\ref{pic:ANN-Error-dist-A}, we show the distribution of $\sigma_{rms}$ over different nuclei from light to heavy. It is clear that on average ANN7 performs the best, but for light nuclei with $A<40$, ANN4 is the best. For nuclei with $120<A<160$, ANN7 and ANN6 work similarly well but ANN6 is slightly better.  The fact that ANN7 is better than ANN6 for light  and heavy nuclei indicates that 
the ASY feature plays an important role. Even though for nuclei with $80\leq A<200$, their RMSDs are  both small, as shown in Fig.~\ref{pic:ANN-Error-N-Z}, the generalizability of ANN7 is  more robust .

It is interesting to study the performance of different network structures as one moves away from the beta stability line.  In Fig.~\ref{pic:ANN-Error-N-Z}, we decompose those nuclei in the test set into seven groups in terms of $|N-Z|$ to judge the predictive powers of ANNs in different isospin-asymmetry regions.It is obvious that ANN7 achieves the most stable and accurate predictions. For nuclei in the $|N-Z|>30$ region, the RMSDs between the predictions of ANN7  and the experimental data stay about only 200 keV.

\begin{figure}[htbp]
  \centering
  \includegraphics[width=15cm]{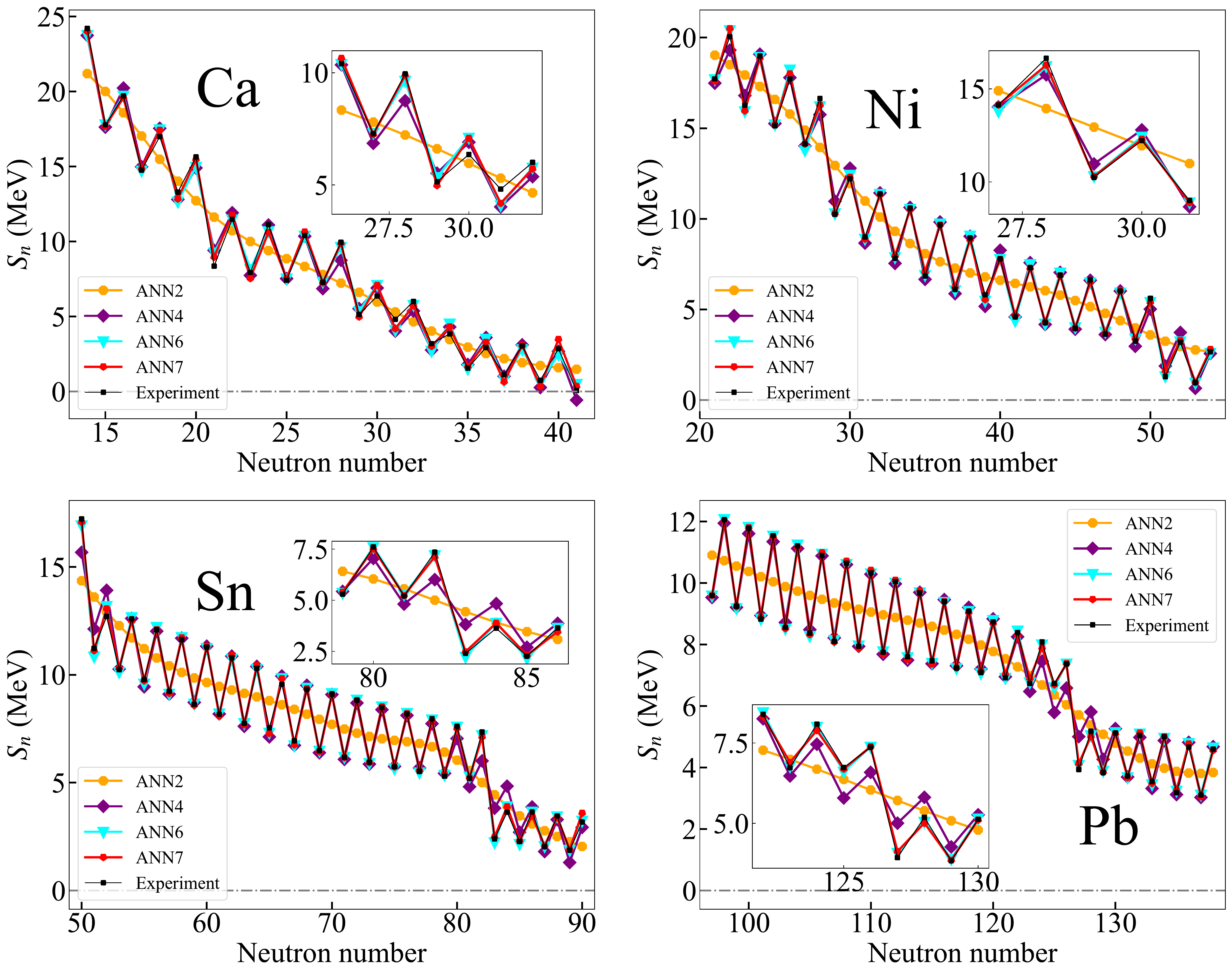}
  \caption{Experimental single neutron separation energies  in comparison with the ANN predictions.}\label{pic:ANN-1n-sep}
\end{figure}

\begin{figure}[htbp]
  \centering
  \includegraphics[width=15cm]{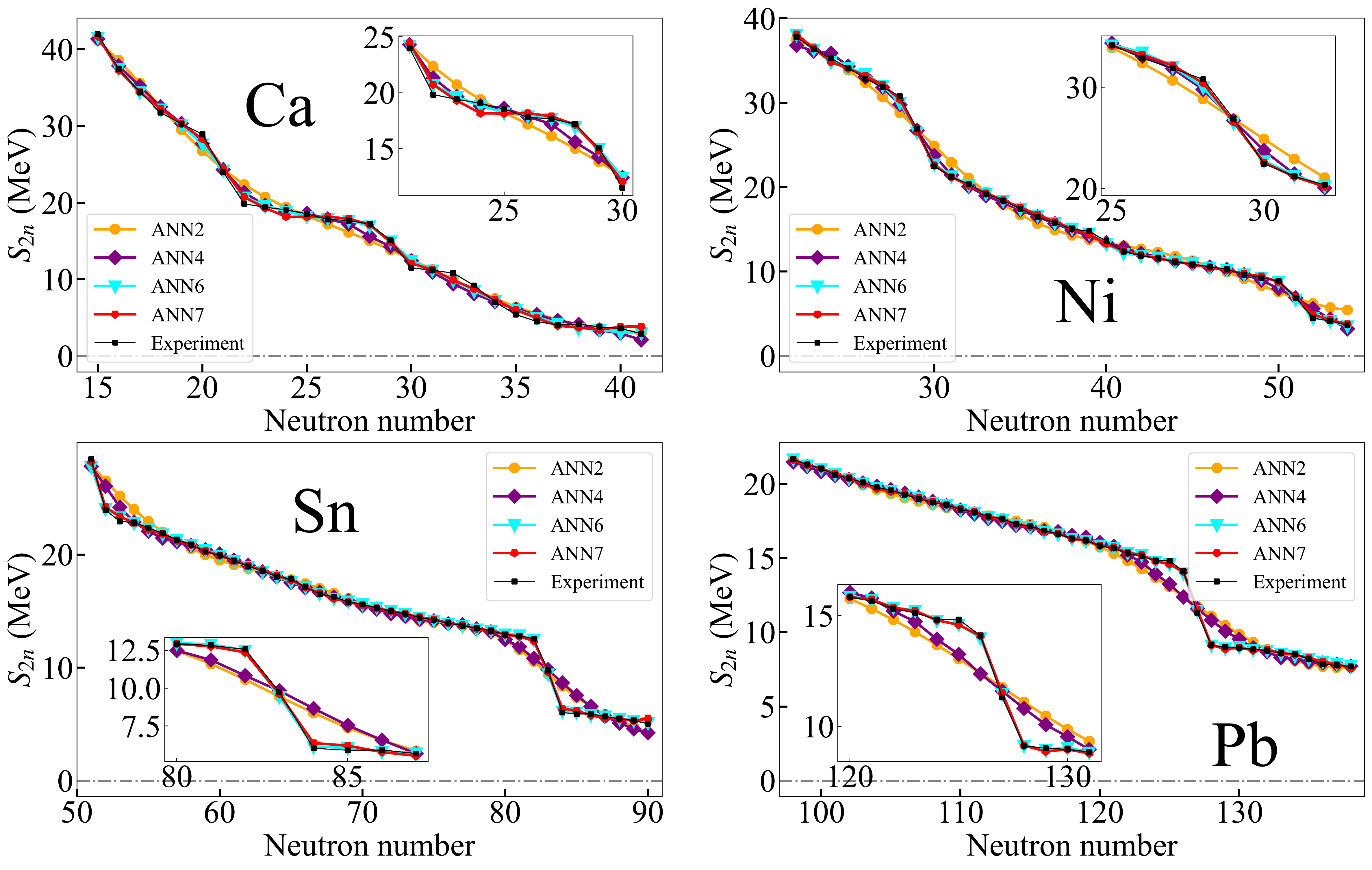}
  \caption{Experimental two-neutron separation energies in comparison with the ANN predictions.}\label{pic:ANN-2n-sep}
\end{figure}

Single and two-neutron separation energies are observables better suited to showcase the details of theoretical models, particularly, whether the shell closure and pairing effects are properly considered. They could be deduced from the binding energies as follows.
\begin{equation}\label{eq:neutron separation energy}
  \begin{array}{rcl}
    S_{n}(Z,N) & = & BE(Z,N)-BE(Z,N-1), \\
    S_{2n}(Z,N) & = & BE(Z,N)-BE(Z,N-2).
  \end{array}
\end{equation}

In Fig.~\ref{pic:ANN-1n-sep} and Fig.~\ref{pic:ANN-2n-sep}, we compare the experimental one-neutron separation energies with the predictions of four ANNs for the Ca, Ni, Sn, and Pb isotopic chains. For $S_{n}$,  ANN2 cannot describe at all the odd-even staggerings, while ANN4 largely improves the situation. However, as is also reflected in Fig.~\ref{pic:ANN-2n-sep}, for nuclei close to the shell closures, the deviations are larger. ANN6 and ANN7, on the other hand, can describe all the nuclei including the  neutron-rich ones.

\begin{figure}[htbp]
  \centering
  \includegraphics[width=15cm]{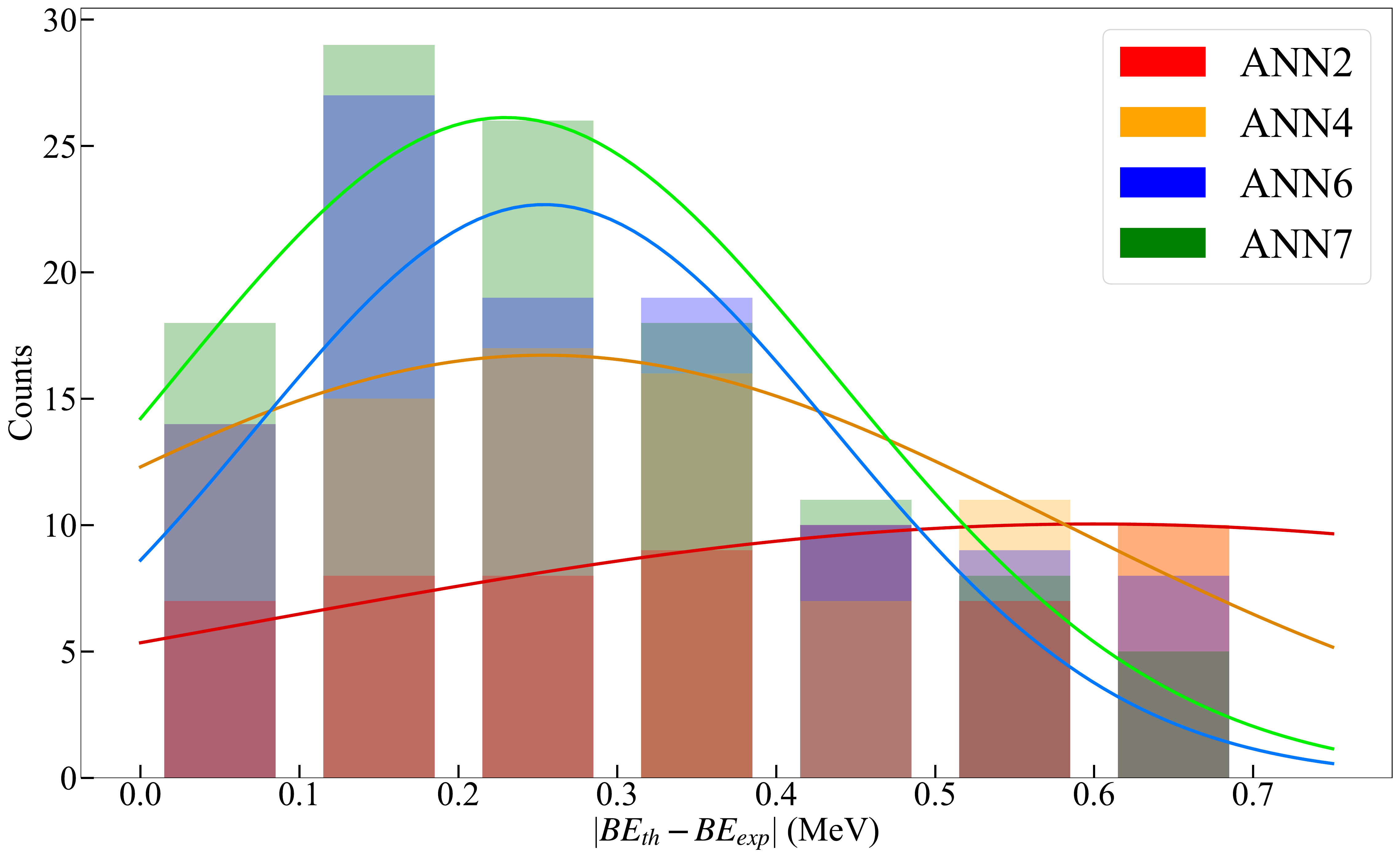}
  \caption{Number of nuclei for which  the BE deviations fall below  0.7 MeV are divided into seven equally spaced groups. The normal distribution curves fit the number of nuclei for which the $\sigma_{rms}$ are lower than 0.7 MeV. We note that  63(51.6\%) of the nuclei in  ANN2, 32(26.2\%) of the nuclei in ANN4, 16(13.1\%) of the nuclei in ANN6 and 7(5.7\%) of the nuclei in ANN7 have  deviations over 0.7 MeV.}\label{pic:ANN-Error-frequency}
\end{figure}

The predictions of a good theoretical model should center around its mean value with small spreads. To check how the four ANNs perform in this perspective,  we  show in Fig.~\ref{pic:ANN-Error-frequency}, the number of nuclei for which the   deviations between theory and experiment fall in between certain ranges. We further fit these counts with normal distributions. The mean of the Gaussian fit indicates the accuracy of the predictions, and the variance reflects the range of deviations. From Fig.~\ref{pic:ANN-Error-frequency}, it could be seen that the Gaussian fits of ANN2 and ANN4 are quite flat, and there are still many deviations that are over 0.7 MeV. In contrast, both ANN6 and ANN7 have very narrow distributions. In addition, most predictions by ANN7 deviate from their experimental counterparts  by less than 0.7 MeV. In this sense, ANN7 not only predicts well but also is more certain.

\subsection{Comparison  with some recent works}

In the past, most machine learning studies of nuclear binding energies adopt the residual approach, which fits the residuals between experimental data and the predictions of an underlying theoretical model~\cite{Carnini:2020lvr,Gao:2021eva,Niu:2018csp,Zhang:2017zvb,Utama:2015hva,WU2022137394}. In the past two years, a number of studies fitting directly to binding energies appeared. In the following, we compare our study with two  recent works.

In Ref.~\cite{Mumpower:2022peg}, with the mixture density network (MDN)~\cite{astonpr373} the authors  used 450 nuclei in AME2016~\cite{wang_ame2016_2017} with $Z\geq20$ as the training set. The first test set  contains all the nuclei  in AME2016 with $Z\geq20$, and the second test set is the latest release of AME2020. The RMSDs of the training set and two test sets are 0.186 MeV, 0.316 MeV, and 0.336 MeV, respectively. We note that although the training sets of ANN7 and MDN are both taken from AME2016, the MDN training set contains fewer nuclei. Light nuclei, which are difficult to describe, are not considered by the MDN model. We note that although the present work and Ref.~\cite{Mumpower:2022peg} adopt different networks, inputs, and  training sets, their performances are rather similar. 

The data augmentation technique, i.e., Gaussian Noise augmentation,  was found in Ref~\cite{Bahtiyar:2022wph} to improve the predictions of ANNs. The number of nuclei in the training set expands from 1685 to 10110. The improvements in different MLP from the perspective of RMSDs are from 18.86\% to 30.50\% in the test set that is in the training data region, and from 23.47\% to 36.33\% in the test set that is beyond the training data region. We also tried to apply the data augmentation technique to  our model, but found that this technique affects little our results.

\section{Summary and outlook}

In this work, we developed a deep neural network with seven physically motivated  features: $Z$, $N$, $Z_{EO}$, $N_{EO}$, $\Delta Z$, $\Delta N$ and ASY. We studied the nuclear masses compiled in AME2020 (measured and extrapolated), achieving a description with a root-mean-square deviation around 0.2 MeV which is much smaller than the previous work~\cite{Yuksel:2021nae} and closer to those of Refs.~\cite{Lovell:2022pkw,Mumpower:2022peg}. The success of our work further demonstrated the importance of considering relevant physical information, i.e., ``feature engineering'', when applying machine learning methods to study systems for which only limited data are available.

It is interesting to note that the description of the nuclear binding energies achieved in the present work is similar to those of Refs.~\cite{Lovell:2022pkw,Mumpower:2022peg} but our work  differs from those of Refs.~\cite{Lovell:2022pkw,Mumpower:2022peg}  in many details: the networks, constraints and input features. In Ref.~\cite{Lovell:2022pkw}, 12 features are used. While in our approach, we found that only six of them ($Z$, $N$, $Z_{EO}$, $N_{EO}$, $\Delta Z$, $\Delta N$) are relevant. On the other hand, Ref.~\cite{Mumpower:2022peg} considered the constraint of the GK relation in addition to eight features. Nevertheless, the similar results achieved in these works support the conclusion that machine learning methods are powerful enough to predict nuclear binding energies at a level comparable to or even better than the most refined conventional theoretical models.

This work reveals that for systems with limited  data, the consideration of input features containing the most relevant physical information can be key to the success for physical studies using machine learning methods. Turning the argument around, by trial and error, one can also anticipate the discovery of ``new physics'' by examining the deficiency of ANNs in describing such systems.

\section{Acknowledgement}

We would like to express our gratitude towards Esra Y$\ddot{u}$ksel and M. R. Mumpower for useful communications.
This work is supported in part by the National Natural Science Foundation of China under Grants No.11735003, No.11975041,  and No.11961141004.

\bibliography{BNN.bib}

\end{document}